\documentclass[aps,prl,twocolumn,superscriptaddress,amsmath,amssymb]{revtex4-1}

\usepackage{hyperref}
\usepackage{graphicx}

\usepackage{amsmath}
\usepackage{bm}
\usepackage{amssymb}
\usepackage{soul}

\usepackage{color}

\begin{document}

\title{Spin-Polarized Electrons in Monolayer MoS$_2$}
\author{Jonas G. Roch}
\email{jonasgael.roch@unibas.ch}
\affiliation{Department of Physics, University of Basel, Klingelbergstrasse 82, CH-4056 Basel, Switzerland}

\author{Guillaume Froehlicher}
\affiliation{Department of Physics, University of Basel, Klingelbergstrasse 82, CH-4056 Basel, Switzerland}

\author{Nadine Leisgang}
\affiliation{Department of Physics, University of Basel, Klingelbergstrasse 82, CH-4056 Basel, Switzerland}

\author{Peter Makk}
\affiliation{Department of Physics, University of Basel, Klingelbergstrasse 82, CH-4056 Basel, Switzerland}
\affiliation{Department of Physics, Budapest University of Technology and Economics and Nanoelectronics ``Momentum" Research Group of the Hungarian Academy of Sciences, Budafoki ut 8, 1111 Budapest, Hungary}

\author{Kenji Watanabe}
\affiliation{National Institute for Material Science, 1-1 Namiki, Tsukuba, 305-0044, Japan}

\author{Takashi Taniguchi}
\affiliation{National Institute for Material Science, 1-1 Namiki, Tsukuba, 305-0044, Japan}

\author{Richard J. Warburton}
\affiliation{Department of Physics, University of Basel, Klingelbergstrasse 82, CH-4056 Basel, Switzerland}

\begin{abstract}
The optical susceptibility is a local, minimally-invasive and spin-selective probe of the ground state of a two-dimensional electron gas. We apply this probe to a gated monolayer of MoS$_2$. We demonstrate that the electrons are spin polarized. Of the four available bands, only two are occupied. These two bands have the same spin but different valley quantum numbers. We argue that strong Coulomb interactions are a key aspect of this spontaneous symmetry breaking. The Bohr radius is so small that even electrons located far apart in phase space interact, facilitating exchange couplings to align the spins.
\end{abstract}

\maketitle

A two dimensional electron gas (2DEG) is formed when the movement of free electrons is limited to two spatial dimensions. As the electron density $n$ increases, single particle effects (phase-space filling leading to a Fermi energy) increases more rapidly than the Coulomb interactions. This ratio is described with the parameter, $r_s=\frac{1}{\sqrt{\pi n}}\frac{1}{a_B}$ where $a_B$ is the effective Bohr radius. Coulomb interactions dominate at large values of $r_s$ where Wigner crystal and ferromagnetic fluid phases are predicted~\cite{Attaccalite2002}. However, in 2DEGs in silicon and gallium arsenide, the electrons are typically localized at large values of $r_s$. 

Monolayer transition metal dichalcogenides (TMDs) such as MoS$_2$, MoSe$_2$, WS$_2$ and WSe$_2$ represent a natural host for a 2DEG. There are two inequivalent conduction band valleys at the $K$ and $K'$ points of the Brillouin zone. The large electron effective mass~\cite{Kormanyos2014} and the weak dielectric screening result in an extremely small Bohr radius, $\sim 0.5$ nm. The immediate consequence is that $r_s$ is pushed towards relatively large values at experimentally relevant electron concentrations. 

MoS$_2$ is a special TMD as the spin-orbit splitting in the conduction band is small compared to typical 2DEG Fermi energies~\cite{GuiBinLiu2014}. There are four available bands: $K_\uparrow$, $K_\downarrow$, $K'_\uparrow$ and $K'_\downarrow$. In a single-particle picture at realistic electron concentrations, the low temperature ground state consists of a close-to-equal filling of the four bands. We present here an experiment which overturns this single-particle picture. We argue that Coulomb correlations are crucial to the explanation.

We probe the 2DEG ground-state in MoS$_2$ by measuring the susceptibility at optical frequencies. This probe is particularly powerful. First, it is a local measurement: signal is gleaned from a few-hundred-nanometer diameter spot on the sample. On this length scale, close-to-ideal optical linewidths have been demonstrated~\cite{Cadiz2017,Ajayi2017} yet there are clearly inhomogeneities on larger length scales even with state-of-the-art material. Second, the measurement represents a weak perturbation to the ground state: in our experiments, there is on average less than one photo-created excitation (an exciton, an electron-hole pair)~(\textit{See Appendix C}). Third, the optical probe is valley- and spin-selective: the electron in the electron-hole pair is created at specific points in the Brillouin zone via the polarization of the light~\cite{DiXiao2012}. Interpreting our results with the established theory of the 2DEG optical susceptibility shows that up to $n \simeq 5 \times 10^{12}$ cm$^{-2}$, two and not four bands are occupied. The electrons have the same spin but reside in different valleys. This is our main result: the spontaneous creation of a spin-polarized 2DEG.

\begin{figure}
\centering
\includegraphics[width=80mm]{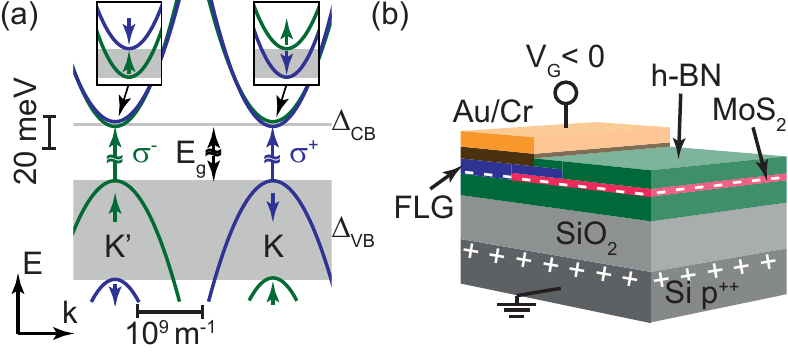}
\caption{(a) Band structure of monolayer MoS$_2$. The color corresponds to the electron spin state. (b) A van der Waals heterostructure consisting of monolayer MoS$_2$  embedded in h-BN all placed on a Si/SiO$_2$ substrate. The MoS$_2$ is contacted by a few-layer graphite (FLG) electrode. The FLG is covered with a Cr/Au layer to which a voltage $V_{\rm G}$ is applied.}
\label{fig1}
\end{figure}

\begin{figure*}
\includegraphics[width=180mm]{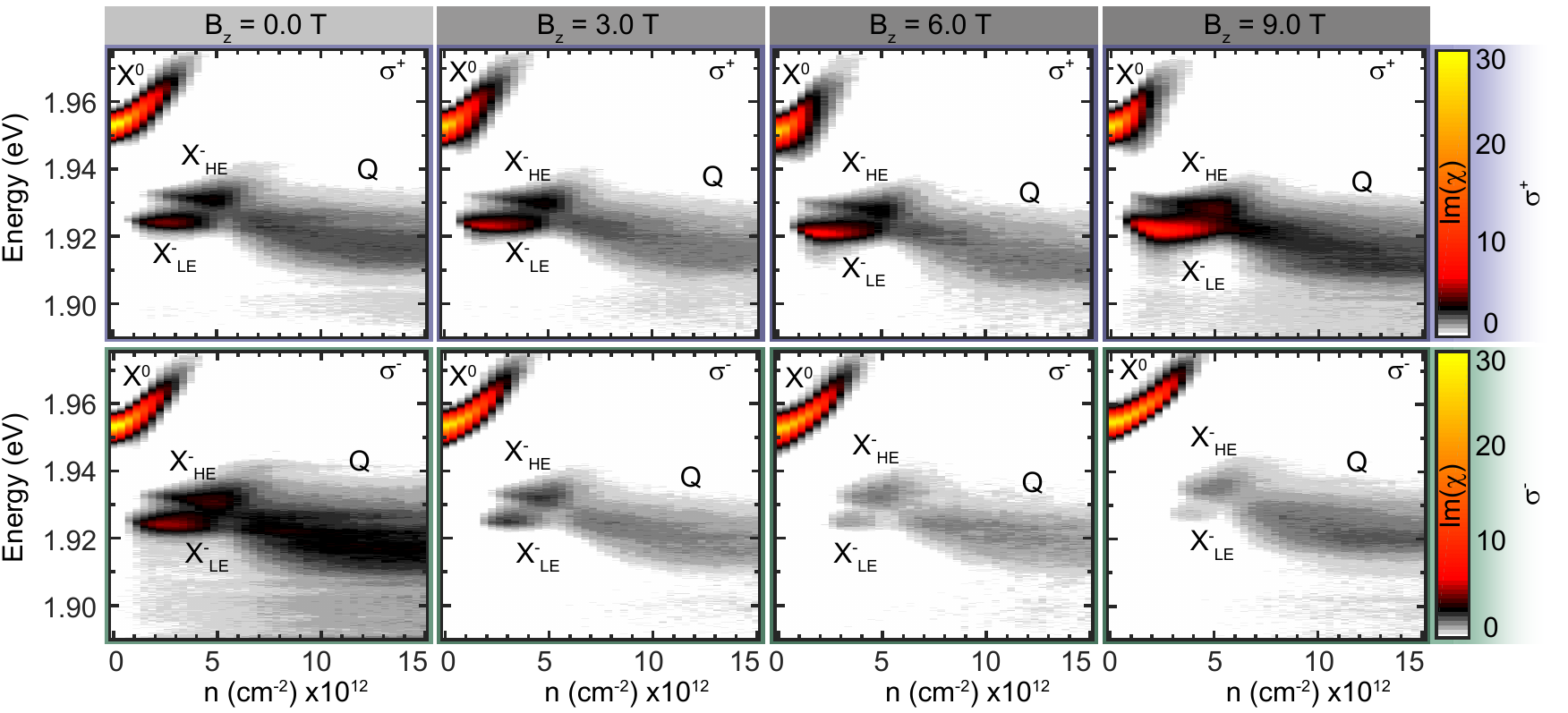}
\caption{Optical susceptibility of a gated MoS$_2$ device as a function of the photon energy (vertical axis) and electron concentration (horizontal axis). The susceptibility is measured in a perpendicular magnetic field $B_z$ of 0.0, 3.0, 6.0 and 9.0 T (colormaps from left to right) with $\sigma^+$-polarized light (top panels) and with $\sigma^-$-polarized light (bottom panels).}
\label{fig2}
\end{figure*}

Monolayer MoS$_2$ has a graphene-like structure with Mo and S sub-lattices~\cite{Mak2010,DiXiao2012}. The band edges are located at the $K$ and $K'$ points; an energy gap of about 2 eV separates the conduction band (CB) from the valence band (VB) (Fig.~\ref{fig1}(a)). Spin degeneracy is lifted by spin-orbit coupling. In each valley, the two CBs are split by $\Delta_{\rm CB} \approx 3$ meV and the the two VBs are split by $\Delta_{\rm VB} \approx 150$ meV~\cite{GuiBinLiu2014}. As Fermi levels of tens of meV are easily achieved, all four CBs are relevant. In MoS$_2$, the upper VB has the same spin as the lower CB~\cite{Mak2013}. Optical absorption promotes an electron from a VB state to a CB state with strict selection rules~\cite{DiXiao2012}: a circularly polarised $\sigma^+$ photon couples the VB and CB with spin-$\downarrow$ at the $K$ point; a $\sigma^-$ photon couples the VB and CB with spin-$\uparrow$ at the $K'$ point (Fig.~\ref{fig1}(a)). 

Fig.~\ref{fig1}(b) shows the structure of our sample. A monolayer of MoS$_2$ forms a planar capacitor with respect to a conductive substrate. The carrier density $n$ in the monolayer is determined by a voltage $V_{\rm G}$ applied to the capacitor: $n=CV_{\rm G}$, where $C\approx 11$ nFcm$^{-2}$ is the geometrical capacitance per unit area~(\textit{See Appendix A}). Efficient injection of electrons into the MoS$_2$ monolayer is ensured by a few-layer graphite (FLG) contact~\cite{Yu2014}. State-of-the-art sample quality and homogeneity is achieved by encapsulation of the MoS$_2$ monolayer in hexagonal boron nitride (h-BN)~\cite{Cadiz2017,SI}.

We measure the optical reflectivity from a 400 nm diameter region on the device at low temperature (4.2 K) using a confocal microscope and ultra-weak, incoherent light source~(\textit{See Appendix B}). The imaginary part of the optical susceptibility is deduced from the reflectivity by accounting for optical interferences~\cite{Back2017,SI}. At the optical resonance, the reflectivity contrast is very large, 60\% (equivalently, the susceptibility is 30), a consequence of the large oscillator strength of the exciton. The first task is to verify the fidelity of the optical selection rules. We do this by tuning $V_{\rm G}$ such that $n \simeq 0$ and by applying a perpendicular magnetic field which splits spectrally the two exciton resonances, one from the $K$ point, the other from the $K'$ point. We observe two distinct resonances, one with $\sigma^+$-polarization, the other with $\sigma^-$-polarization, with no cross-talk within the noise~(\textit{See Appendix D}). This demonstrates the optical spin-valley effect, equivalently the high spatial homogeneity of our sample.

\begin{figure*}
\includegraphics[width=180mm]{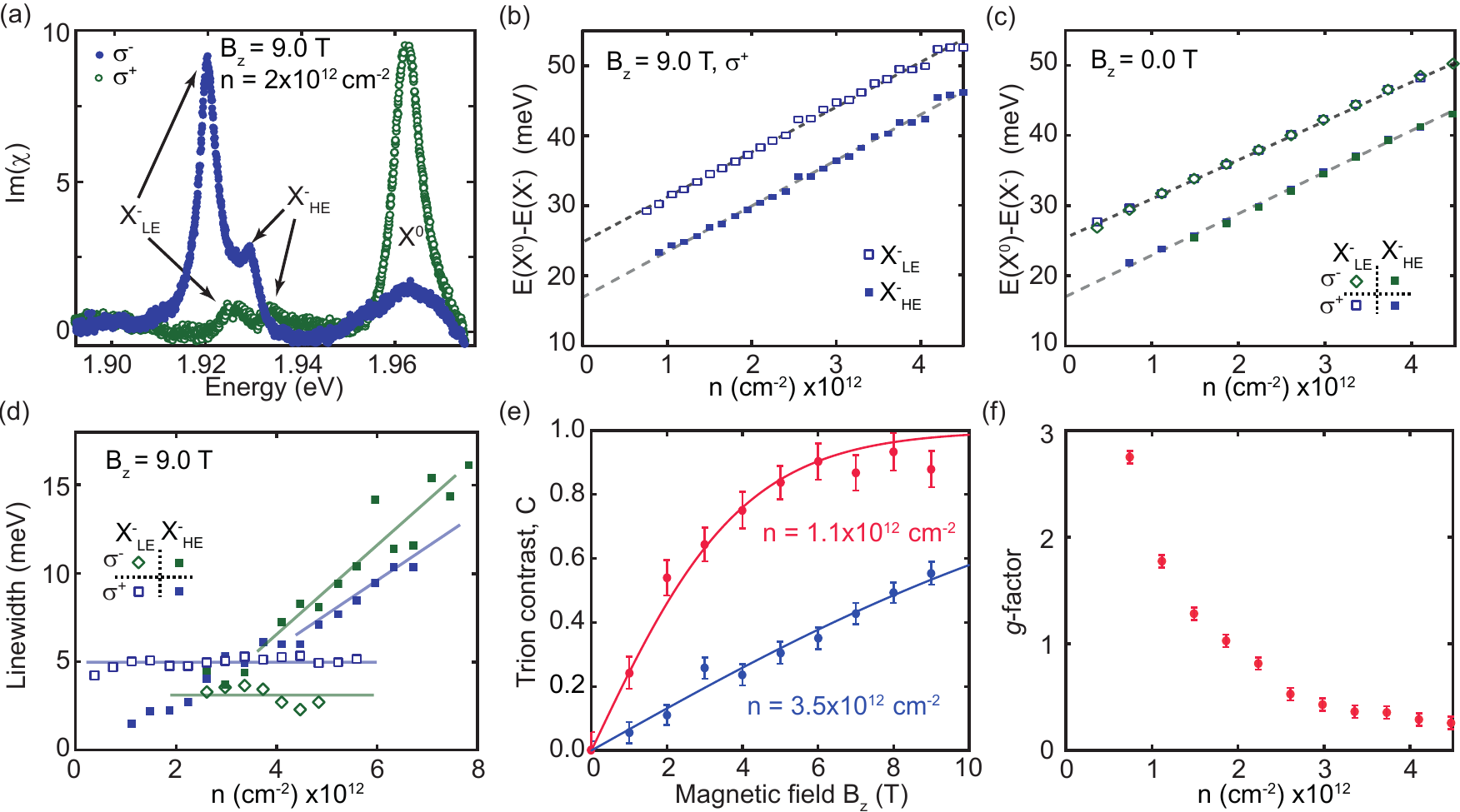}
\caption{(a) Optical susceptibility at $B_z = 9.0$ T and $n=2.0 \times 10^{12}$ cm$^{-2}$ for both $\sigma^+$ and $\sigma^-$ photons. For $\sigma^+$, two trion resonances ($X^-_{\rm LE}$ and $X^-_{\rm HE}$) dominate the spectrum. For $\sigma^-$, $X^0$ dominates and $X^0$ has a high energy tail. (b) Energetic difference $E(X^0)-E(X^-)$ for the two trions at $B_z=9.0$ T as a function of $n$. Linear fits give slopes of $6.1\times 10^{-15}$ and  $6.3\times 10^{-15}$ eVcm$^2$ for $X^-_{\rm LE}$ and $X^-_{\rm HE}$, respectively. (c) As in (b), but at $B_z = 0.0$ T. Linear fits give $5.6\times 10^{-15}$ and  $5.9\times 10^{-15}$ eVcm$^2$ for $X^-_{\rm LE}$ and $X^-_{\rm HE}$, respectively. The trion binding energies are 17~meV ($X^-_{\rm LE}$) and 25~meV ($X^-_{\rm HE}$). (d) Linewidth of $X^-_{\rm LE}$ and $X^-_{\rm HE}$ at $B_z=9.0$ T versus $n$. (e) Contrast  $C$ (as defined in the text) versus $B_z$ for $n=1.1\times 10^{12}$~cm$^{-2}$ (red) and for $n=3.7\times 10^{12}$~cm$^{-2}$ (blue). The solid lines are a fit to two-level Maxwell-Boltzmann statistics with notional $g$-factor $g=1.6 \pm 0.1$ (red) and $g=0.4 \pm 0.1$ (blue). (f) Notional $g$-factor as a function of $n$. Taking a Bohr radius of 0.48~nm, $n=1.0\times 10^{12}$~cm$^{-2}$ corresponds to $r_s=11.8$. }
\label{fig3}
\end{figure*}

Fig.~\ref{fig2} shows as colormaps the optical susceptibility as a function of electron concentration at various magnetic fields $B_z$ applied perpendicular to the 2DEG plane. The second task is to identify the spectral resonances. We focus initially on the susceptibility at $B_z=9.0$ T.

At low electron density, the peak labeled $X^0$ dominates the susceptibility for both $\sigma^+$ and $\sigma^{-}$. This resonance corresponds to the creation of a Coulomb-bound electron-hole pair, the neutral exciton. As $n$ increases, the $X^0$ blue-shifts, broadens and weakens, eventually disappearing into the noise. With $\sigma^+$-polarization, as $X^0$ weakens, two resonances emerge, labelled $X^-_{\rm LE}$ and $X^-_{\rm HE}$. These two resonances are red-shifted with respect to $X^0$. In the opposite photon polarization, $\sigma^-$, as $n$ increases the $X^0$ resonance blue-shifts and weakens but for low to modest $n$ there are no $X^-$ features.

The excitons injected into the 2DEG by our optical probe interact with the electrons in the Fermi sea. Theory has been developed to describe an exciton interacting either with a spin-degenerate Fermi sea at the $\Gamma$-point of the Brillouin zone~\cite{Suris2001} or a spin-polarized Fermi sea~\cite{Efimkin2017}, and successfully stress-tested against experiments on quantum wells~\cite{Suris2001}. When the exciton-electron interaction is attractive, the exciton resonance splits into two corresponding to the formation of so-called exciton-polarons~(\textit{See Appendix E}). The upper exciton-polaron corresponds to $X^{0}$; the lower exciton-polaron corresponds to $X^-$ (and becomes the trion in the single-particle limit~\cite{Mak2013,Plechinger2016,Courtade2017}). On the other hand, when the exciton-electron interaction is repulsive, only the $X^0$ appears in the susceptibility with a tail on the high energy side~(\textit{See Appendix E}). In these theories, the interaction is attractive only if the electron in the exciton and the electron in the Fermi sea have opposite spins. This is an electron spin-singlet. The interaction is repulsive for parallel spins, the spin-triplet.

We apply the exciton-polaron theory to the MoS$_2$ susceptibility. With $\sigma^+$-photons, we observe not one but \textit{two} lower exciton-polarons, $X^-_{\rm LE}$ and $X^-_{\rm HE}$. This implies that the exciton interacts with \textit{two} Fermi seas, the energy splitting arising from a different exciton-electron scattering cross-section. Specifically, the electron in the exciton has spin-$\downarrow$. To form spin-singlets, both Fermi seas must have spin-$\uparrow$, i.e.\ the $K'_\uparrow$ and $K_\uparrow$ bands are occupied. The different binding energies (defining the binding energy as the energy separation between the two exciton-polarons in the limit $n \rightarrow 0$) arise from the fact that the two Fermi seas are at different locations in phase-space. The $X^-_{\rm LE}$ has a constant linewidth whereas the $X^-_{\rm HE}$ has a linewidth which increases with $n$ (Fig.~\ref{fig3}(d)). This difference also points to the fact that the exciton interacts with two different Fermi seas. With $\sigma^-$-photons, the spectra follow the exciton-polaron theory for a repulsive exciton-Fermi sea interaction. This means that the photo-excited spin-$\uparrow$ electrons interacts with spin-$\uparrow$ electrons in the Fermi sea: there are no spin-$\downarrow$ partners to create spin-singlets in this case. For both $\sigma^+$- and $\sigma^-$-photons, the details of the measured spectra match the exciton-polaron theory~(\textit{See Appendix E}).

\begin{figure*}
\includegraphics[width=180mm]{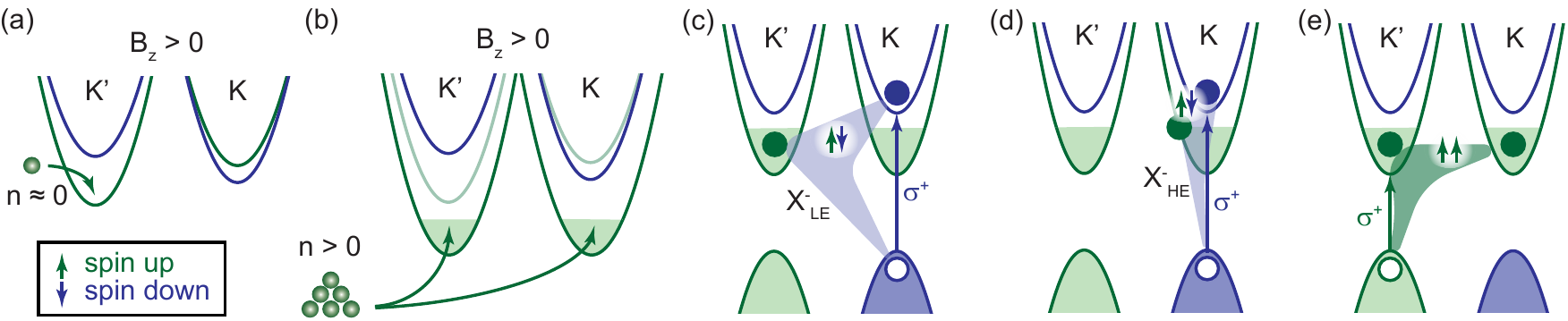}
\caption{(a) The conduction bands at $B_z=9$ T. The minima all have different energies due to Zeeman shifts~(\textit{See Appendix D}). The ``first" electrons injected from the contact populate the band with the lowest energy, $K'_{\uparrow}$. (b) Strong intra- and inter-valley exchange pull the bands of the same spin to lower energy, creating a fully spin-polarized 2DEG. (c) Exciton-Fermi sea interaction resulting in $X^-_{\rm LE}$. (d) Exciton-Fermi sea interaction resulting in $X^-_{\rm HE}$. The trions in (c) and (d) both correspond to electron spin-singlets. (e) A trion with an electron spin-triplet. This trion resonance is unbound and results in the high energy tail of the $X^0$ with $\sigma^-$ photons.}
\label{fig4}
\end{figure*}

Interpreting the optical susceptibility spectra with the established exciton-polaron theory therefore leads us to the conclusion that at $B_z=9.0$ T, two bands are occupied, both with spin-$\uparrow$. Fig.~\ref{fig4}(c-d) show how we understand the two $X^-$ resonances. A photon-generated electron-hole pair with electron spin-$\downarrow$ interacts attractively with spin-$\uparrow$ electrons from two different bands. The trion binding for the inter-valley scattering process (Fig.~\ref{fig4}(c)) is larger~\cite{Yu2014v2,Yu2015} as the electrons have both opposite spin and valley indices, similar to MoSe$_2$~\cite{Back2017}. We associate this process to the resonance $X^-_{\rm LE}$. The intra-valley scattering process (Fig.~\ref{fig4}(d)) leads to the resonance $X^-_{\rm HE}$. The absence of an $X^-$ resonance in $\sigma^-$ polarisation tells us that the triplet process in Fig.~\ref{fig4}(e) is unbound.

Of the four bands, only \textit{two} are occupied. This conclusion on the number of occupied bands can be verified via another feature of the susceptibility spectra. In the limit of large hole mass, the energetic separation between the upper and lower exciton-polarons is simply $\delta E=E(X^0)-E(X^-)=E_{\rm B}+E_{\rm F}$ where $E_{\rm B}$ is the trion binding energy and $E_{\rm F}$ the Fermi energy~\cite{Hawrylak1991}, a result demonstrated experimentally on CdTe quantum wells~\cite{Huard2000}. For the equal hole and electron masses of MoS$_2$, this result applies at $E_{\rm F} \geq 20$ meV~(\textit{See Appendix E}). The gradient $\mathrm{d}\delta E/\mathrm{d}E_{\rm F}$ increases from 1.0 to 2.0 as $E_{\rm F} \rightarrow 0$. Our experimental data lie mostly in the high-$E_{\rm F}$ regime~(\textit{See Appendix E}). This enables us to determine $E_{\rm F}$ from the optical spectra. As $E_F$ is linked to $n$ by the two-dimensional density of states, we can determine how many bands are populated. Taking an electron effective mass of $m^{*}_{e}=0.44 m_{o}$~\cite{Kormanyos2014}, the measured $\mathrm{d}\delta E/\mathrm{d}n$ (Fig.~\ref{fig3}(b)) implies that $1.9\pm 0.1$ bands are occupied at $B_z=9.0$ T~(\textit{See Appendix F}).

We turn now to the magnetic field dependence for $n \leq 6 \times 10^{12}$ cm$^{-2}$. At $B_z=0.0$ T, the $X^-_{\rm LE}$ and $X^-_{\rm HE}$ features are equally strong for both $\sigma^+$ and $\sigma^-$ photons. As $B_z$ increases, the $X^-_{\rm LE}$ and $X^-_{\rm HE}$ gradually disappear for $\sigma^-$ photons. At $B_z=0.0$ T, the gradients $\mathrm{d}\delta E/\mathrm{d}n$ (Fig.~\ref{fig3}(c)) change by less than 10\% with respect to $B_z=9.0$ T, suggesting that even in this limit, only two bands are occupied. We define a contrast $C$ as
\begin{equation}
C=\frac{I_{\rm LE+HE}(\sigma^+) - I_{\rm LE+HE}(\sigma^-)}{I_{\rm LE+HE}(\sigma^+) + I_{\rm LE+HE}(\sigma^-)},
\end{equation}
where $I_{\rm LE+HE}(\sigma^+)$ [$I_{\rm LE+HE}(\sigma^-)$] is the integrated susceptibility of $X^-_{\rm LE}$ and $X^-_{\rm HE}$ in $\sigma^+$ [$\sigma^-$] polarisation. In the most extreme case, $C$ increases from $C=0$\% at $B_z=0.0$ T to $C=95$\% at $B_z=9.0$ T (Fig.~\ref{fig3}(e)). Phenomenologically, we imagine that flipping an electron spin costs an energy $g \mu_B B_z$ where $\mu_B$ is the Bohr magneton and $g$ is a $g$-factor. Maxwell-Boltzmann statistics applied to this notional two-level system gives $C(B)=\tanh(g \mu_B B_z/k_{\rm B} T)$, where $k_{\rm B}$ is the Boltzmann constant and $T$ the temperature. $C$ follows this dependence on $B$ (Fig.~\ref{fig3}(e)). In fact $g$ can be extracted at different values of $n$ (Fig.~\ref{fig3}(f)): we find that $g$ decreases strongly with increasing $n$ (decreasing $r_s$), becoming small at high $n$. This dependence on $n$ is entirely characteristic of Coulomb effects: this feature of the experiment suggests strongly that the spin polarization arises as a consequence of Coulomb correlations. 

For $n \geq 6 \times 10^{12}$ cm$^{-2}$, the optical response changes: $X^-_{\rm LE}$ and $X^-_{\rm HE}$ weaken and the susceptibility is dominated by a broad, red-shifted peak labeled $Q$ (Fig.~\ref{fig2}). There is no established theory for the optical susceptibility in this regime where the Fermi energy exceeds the trion binding energy. It is therefore challenging to make definitive statements in this high-$n$ regime. Nevertheless, we speculate that the absence of a marked contrast between $\sigma^+$ and $\sigma^-$ photons signals that the 2DEG is no longer spin-polarized. 

The spin-polarization of the MoS$_2$ 2DEG can be qualitatively understood by exchange~\cite{Scrace2015} and the strong inter-valley Coulomb scattering~\cite{Dery2016}. Fig.~\ref{fig4}(a) depicts the four CBs in a magnetic field. At low temperature, the ``first" injected electrons populate the band with lowest energy. Intra-valley and inter-valley exchange will then favor population of the bands with the same electron spin, as sketched in Fig.~\ref{fig4}(b). The small CB spin-orbit splitting allows occupation of the higher CB with a moderate cost in kinetic energy ($\Delta_{CB}\approx 3$ meV). These results highlight a very particular feature of TMDs. The Bohr radius is only slightly larger than the lattice constant such that the two-body Coulomb interaction connecting an electron at the $K$ point with an electron at the $K'$ point (far apart in phase space) is comparable to the two-body Coulomb interaction between two electrons close together in phase space~\cite{Dery2016}. This equivalence of intra-valley and inter-valley Coulomb interactions is a striking feature. 

At first sight, the spin polarization mimics Stoner ferromagnetism. However, the Stoner mechanism is based on a mean-field theory which is invalid in two-dimensions where ferromagnetic order is excluded by the Mermin-Wagner theorem for any finite temperature~\cite{Mermin1966}. The conduction band spin-orbit splitting, small but non-zero, does however establish an in-built quantization axis such that a spontaneous symmetry breaking is feasible, Mermin-Wagner notwithstanding, but theory hinges on complex diagrammatic techniques \cite{Zak2010}.

\begin{acknowledgments}
We thank Jelena Klinovaja, Daniel Loss and Dmitry Miserev, also Guido Burkard, Alexander Pearce and Andor Korm\'anyos, for fruitful discussions. The work in Basel was financially supported by SNF (Project No.\ 200020\_156637), NCCR QSIT and QCQT. PM acknowledges support from grants OTKA PD-121052, OTKA FK123894 and the Bolyai Fellowship. Growth of hexagonal boron nitride was supported by the Elemental Strategy Initiative conducted by the MEXT Japan and JSPS KAKENHI Grant Numbers JP26248061, JP15K21722, JP25106006.
\end{acknowledgments}

\appendix

\begin{figure*}
\centering
\includegraphics[width=180mm]{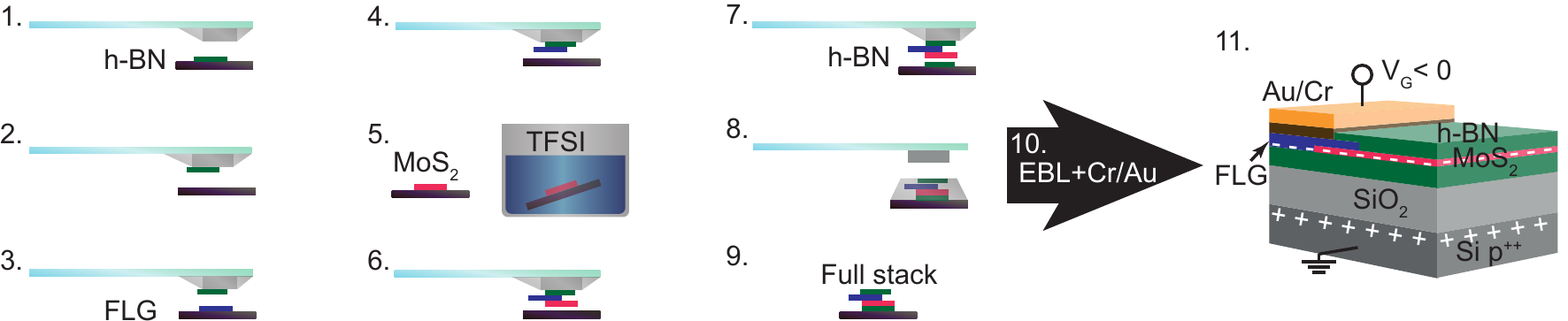}
\caption{\textbf{Sample fabrication.} Using a PDMS/PC stamping procedure, flakes are picked up (1-7) and stacked on the PC film. At the last step (8), the PC film is left on a SiO$_2$/Si substrate by heating. After dissolving the PC film (9), electron beam lithography (EBL) and contact deposition can be performed (10). The final sample structure is sketched in (11).}\label{sampleFab}
\end{figure*}

\subsection*{Appendix A: Sample fabrication}
Van der Waals heterostructures were fabricated by stacking two-dimensional materials via a dry-transfer technique\cite{Zomer2014}, as depicted in Fig.~\ref{sampleFab}. A polydimethylsiloxane (PDMS) stamp with a thin polycarbonate (PC) layer is used to pick up flakes exfoliated on SiO$_2$(300~nm)/Si substrates. Exfoliation was carried out from bulk crystals (natural MoS$_2$ crystal from SPI Supplies, synthetic h-BN\cite{Taniguchi2007}, and natural graphite from NGS Naturgraphit). MoS$_2$ monolayers were treated by a bis(trifluoromethane)sulfonimide (TFSI) solution, following Ref. \cite{Amani2015},  before full encapsulation between h-BN layers. Few-layer graphene (FLG) was employed as a contact electrode to MoS$_2$\cite{Yu2014}. Metal contacts to the FLG were patterned by electron-beam lithography (EBL) and subsequent metal deposition of Au (45~nm)/Cr (5~nm).\\

The capacitance of the device was estimated using electrostatics: the MoS$_2$ flake and the p-doped silicon are considered as two electrodes separated by the thickness $d_\mathrm{BN}$ of the bottom h-BN and the thickness $d_{\mathrm{SiO_2}}$ of the SiO$_2$. The capacitance per unit area is then given by
\begin{equation}
C=\dfrac{1}{\dfrac{d_{\mathrm{BN}}}{\epsilon_{\mathrm{BN}}}+\dfrac{d_\mathrm{{SiO_2}}}{\epsilon_\mathrm{{SiO_2}}}}~~ ,\label{capa}
\end{equation}
where $\epsilon_\mathrm{BN}=3.76$\cite{Laturia2018} and $\epsilon_\mathrm{SiO_2}=3.9$ are the dielectric constants of h-BN and SiO$_2$, respectively. Using $d_\mathrm{BN}=10$~nm and $d_\mathrm{SiO_2}=300$~nm, we obtain $C=11.1\pm 0.5$~nFcm$^{-2}$, where an overall 5\% uncertainty in the layers thicknesses is taken into account. 

Due to the small size of the sample ($\sim 10$~$\mu$m$^2$), it is difficult to measure directly its absolute $\sim$fF capacitance. However, prior to measurements, we checked at room temperature for the absence of a gate leak. No gate leak ($>10~$G$\Omega$) was observed up to 100~V between the top electrode and the bottom gate.\\

Experimentally, we apply a voltage $V_G$ to inject carriers in the 2DEG. For a capacitive device, the carrier concentration $n$ is given by $n=CV_G$ and it is expected that $n=0$ when $V_G=0$. However, as a combined consequence of photo-doping effect~\cite{FabJu2014,Epping2017} and charge trapping~\cite{Wang2010} at the different interfaces in the van der Waals heterostructure, the device exhibits hysteresis when $V_G$ is swept in a loop. Optically, the absence of negatively charged excitons in the absorption spectra in both polarisation can be used to attest the absence of electrons in the 2DEG. In our device, $n\approx0$ when $V_G(n\approx 0)=100$~V. At different gate voltages, $n$ reads
\begin{equation}
n(V_G)=-C\left(V_G(n\approx 0)-V_G\right).\label{n}
\end{equation}
In the main text, we use Eq.~\ref{n} to determine the electron density $n$.

\subsection{Appendix B: Experimental setup}
\begin{figure}
\centering
\includegraphics[width=80mm]{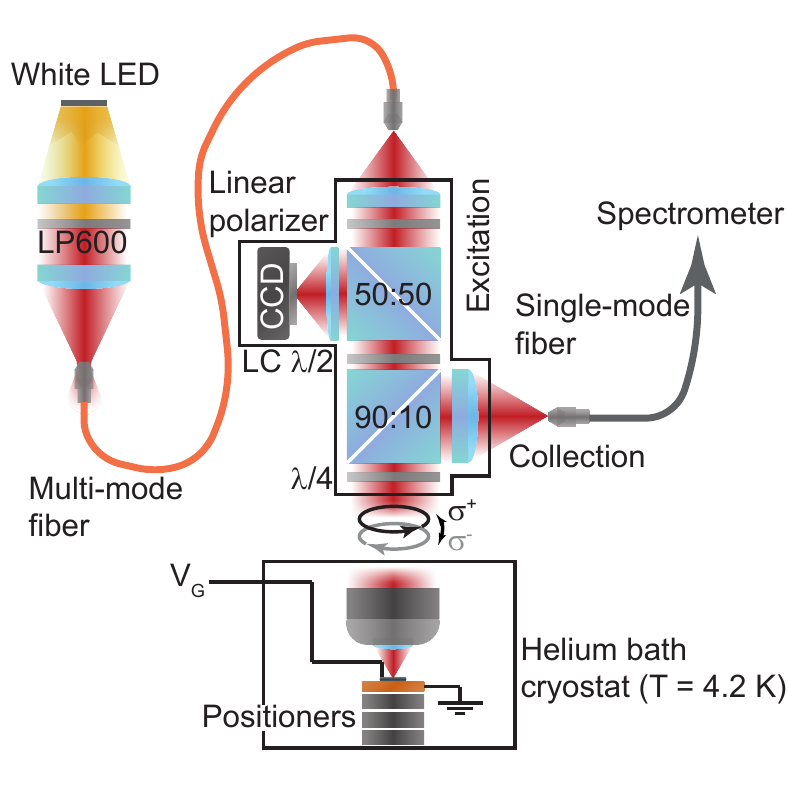}\caption{\textbf{Experimental setup.}}\label{expSetup}
\end{figure}

The absorption spectra diplayed in Fig.~2 in the main text were recorded with the setup sketched in Fig.~\ref{expSetup}. The red part of a white (Osram warm white) light emitting diode (LED) is filtered by a 600~nm longpass filter and coupled into a multi-mode fiber. The output of the fiber is connected to a home-built microscope. A CCD camera mounted on the microscope is used to visualise the sample. In the microscope, the light is first sent through a linear polarizer. A computer controlled liquid crystal (LC) retarder that can produce a $+\lambda/2$ or $-\lambda/2$ retardance of the initial beam is used to produce two perpendicular linear polarizations on demand. An achromatic quarter-wave plate retarder ($\lambda/4$) is subsequently used to produce circularly polarized light. By controlling the voltage on the LC retarder, we can then circularly polarize the LED light with right- or left- handed orientation.
The circularly polarized light goes then in a helium bath cryostat at 4.2~K  and is focused on the sample using a microscope objective (NA=0.65). The position of the sample with respect to the focus can be adjusted with cryogenic nanopositionners. The reflected light  is coupled into a single-mode fiber, ensuring a confocal detection, and sent to a spectrometer. Ligth was dispersed by a 1500 grooves per millimeter grating, before being focused onto a liquid-nitrogen cooled charged coupled device (CCD) array. The spectral resolution of the spectrometer setup is 0.05~nm. In this way, reflectivity spectra were acquired.

\subsection*{Appendix C: Reflectivity of a thin film: determination of the susceptibility}
Monolayers MoS$_2$ or h-BN are visible when placed on a SiO$_2$ substrate. Optical contrast is induced by interferences in the thin films.  Thin film interferences thus play an important role when measuring the reflectivity of a van der Waals heterostructure.

\begin{figure*}
\centering
\includegraphics[width=18cm]{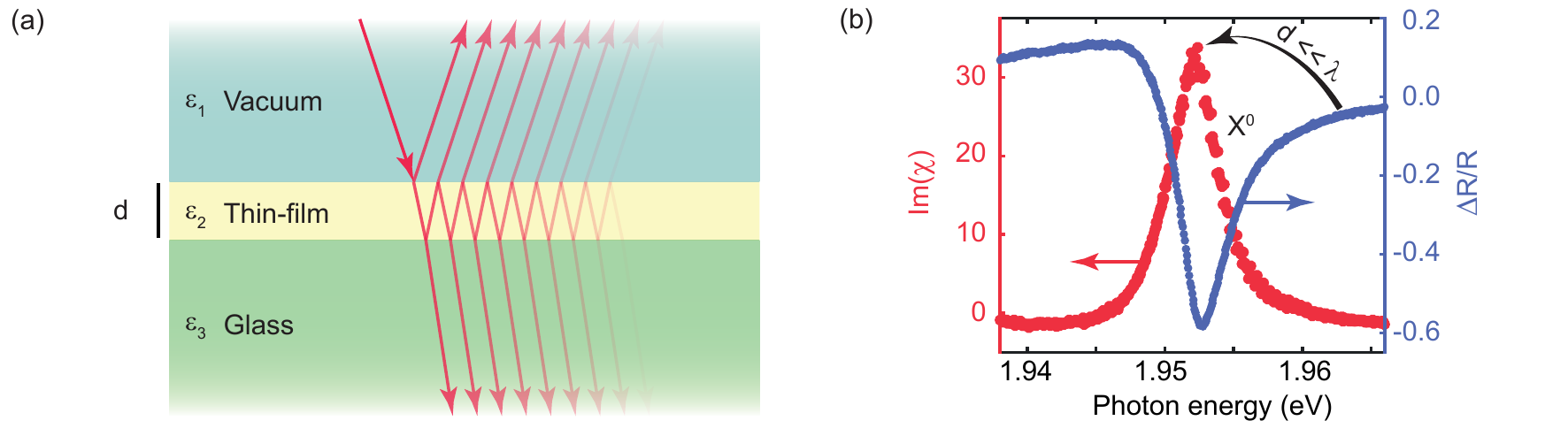}
\caption{\textbf{From reflectivity to the optical susceptibility.} (a) Three layer system: light comes from a semi-infinite medium 1 (top) and is reflected at the thin-film (middle). Multiple interferences in the thin-film create optical interferences. (b) Using a Kramers-Kronig relation, the raw reflectivity data (blue) around the $X^0$ energy at zero electron concentration can be turned into a Lorentzian absorption lineshape. Here, we used a phase factor $\zeta=0.69$~rad to account for multiple thin-film interferences.}
\label{pic3Phases}
\end{figure*}

For a monolayer placed on a thick substrate, the system can be modeled by a three layer system, as depicted in Fig.~\ref{pic3Phases}(a). A thin-film is at the interface between two semi-infinite systems, namely vacuum and the SiO$_2$ layer. As the monolayer thickness $d$ is much smaller than the wavelength of light $\lambda$, the differential reflectivity of the monolayer $\frac{\Delta R}{R}$ at normal incidence can be written as \cite{McIntyre1971}:
\begin{equation}
\frac{\Delta R}{R}=-\frac{8\pi d n_1}{\lambda}\mathbb{I}m\left(\frac{\epsilon_1-\tilde{\epsilon}_2}{\epsilon_1-\epsilon_3}\right), \label{eqMcIntyre}
\end{equation}
where $\epsilon_j$ is the dielectric function of the $j$-th layer and $n_1$ is the refractive index of the first medium. The meaning of the tilde in $\tilde{\epsilon}_2$ will be explained later. In the linear response approximation, $\epsilon_j=1+\chi_j$, where $\chi_j$ is the optical susceptibility of larger $j$. As the first medium is air, $\epsilon_1=1$ and the numerator in Eq.~\ref{eqMcIntyre} simplifies to $\epsilon_1-\tilde{\epsilon}_2=1-(1+\tilde{\chi}_2)=-\tilde{\chi}_2$. Assuming that the glass substrate has a negligible absorption ($\mathbb{I}m(\epsilon_3)=0$), Eq.~\ref{eqMcIntyre} can then be rewritten as, 
\begin{equation}
\mathbb{I}m(\tilde{\chi}_2)=\underbrace{\frac{\lambda (1-\epsilon_3)}{8\pi d}}_{\beta}\frac{\Delta R}{R}\approx -49.1\frac{\Delta R}{R}~~,
\label{realUnits}
\end{equation}
where we use an effective monolayer thickness $d=0.65$~nm, $\epsilon_3=2.25$ and a constant wavelength $\lambda=642$~nm, corresponding to to the centre position of the spectrometer grating during the measurements. Eq.~\ref{realUnits} allows a measured $\frac{\Delta R}{R}$ to be converted into the imaginary part of the susceptibility. We define here $\beta=\frac{\lambda (1-\epsilon_3)}{8\pi d}\approx -49.1$.

In our sample, we have more than three layers. In the experiment, when we measure $\frac{\Delta R}{R}$, we actually probe an effective susceptibility  $\tilde{\chi_2}$ of the MoS$_2$ monolayer. Reflection on the SiO$_2$/Si interface, as well as the multiple thin-film interferences in the h-BN layers will indeed mix the real and imaginary part of the dielectric function of the MoS$_2$. 

As the susceptibility is a complex number, a phase factor $e^{i\zeta}$ can be used\cite{Arora2013,Back2017} to mix the imaginary and real part of the dielectric function of the monolayer MoS$_2$. The effective susceptibility   $\tilde{\chi_2}$ of the thin-film that we probe in our measurement is then related to the susceptibility $\chi_2$ of MoS$_2$ by $\tilde{\chi}_2=e^{-i\zeta}\chi_2$.

We are not interested in the value of the effective susceptibility, but we want to access to the value of $\chi_2$, which is intrinsic to MoS$_2$. However, Eq.~\ref{realUnits} links the measured $\frac{\Delta R}{R}$ to the effective optical susceptibility 
\begin{equation}
\frac{\Delta R}{R}=\frac{1}{\beta}\mathbb{I}m(\tilde{\chi_2})=\frac{1}{\beta}\tilde{\chi_2}''~~,
\end{equation}
where we decomposed $\tilde{\chi_2}$ in its real and imaginary part $\tilde{\chi_2}=\tilde{\chi_2}'+i\tilde{\chi_2}''$. As before, $\beta\approx -49.1$. By definition, $\chi_2=e^{i\zeta}\tilde{\chi_2}$ and therefore 
\begin{equation}
\chi_2''=\cos(\zeta)\tilde{\chi_2}''+\sin(\zeta)\tilde{\chi_2}'~.\label{phaseFac}
\end{equation}

Our experiment measures $\frac{\Delta R}{R}=\frac{1}{\beta}\tilde{\chi_2}''$. In order to compute $\tilde{\chi_2}''$ from Eq.~\ref{phaseFac}, we need to know the value of $\tilde{\chi_2}'$. The causality of the dielectric function $\tilde{\chi_2}$ implies that we can make use of the Kramers-Kronig relation in Eq.~\ref{phaseFac}.
\begin{equation}
\chi_2''(\omega) =\cos(\zeta)\tilde{\chi_2}''+\sin(\zeta)\underbrace{\dfrac{2}{\pi}\mathcal{P}\int\limits_0^\infty \dfrac{\omega'\tilde{\chi_2}''(\omega')}{\omega'^2-\omega^2}\mathrm{d}\omega'}_{\tilde{\chi_2}'}~~,
\end{equation}
where $\omega$ denotes the angular frequency. Using Eq.~\ref{realUnits}, we can express the imaginary part of the dielectric function of MoS$_2$ as a function of the differential reflectivity $\frac{\Delta R}{R}$ that we measure experimentally:
\begin{equation}
\chi_2''=\beta\left(\sin(\zeta)\dfrac{2}{\pi}\mathcal{P}\int\limits_0^\infty \dfrac{\omega'\frac{\Delta R}{R}(\omega')}{\omega'^2-\omega^2}\mathrm{d}\omega'+\cos(\zeta)\frac{\Delta R}{R}\right)~.\label{theWhole}
\end{equation}

As interference effects depend on the wavelength, the phase $e^{i\zeta}$ also depend on the wavelength ($\zeta=\zeta(\omega)$). However, as the wavelength range we access in a single spectrum is small compared to the wavelength itself, to first order, we can set $\zeta$ to a constant value.\\

\begin{figure*}
\centering
\includegraphics[width=180mm]{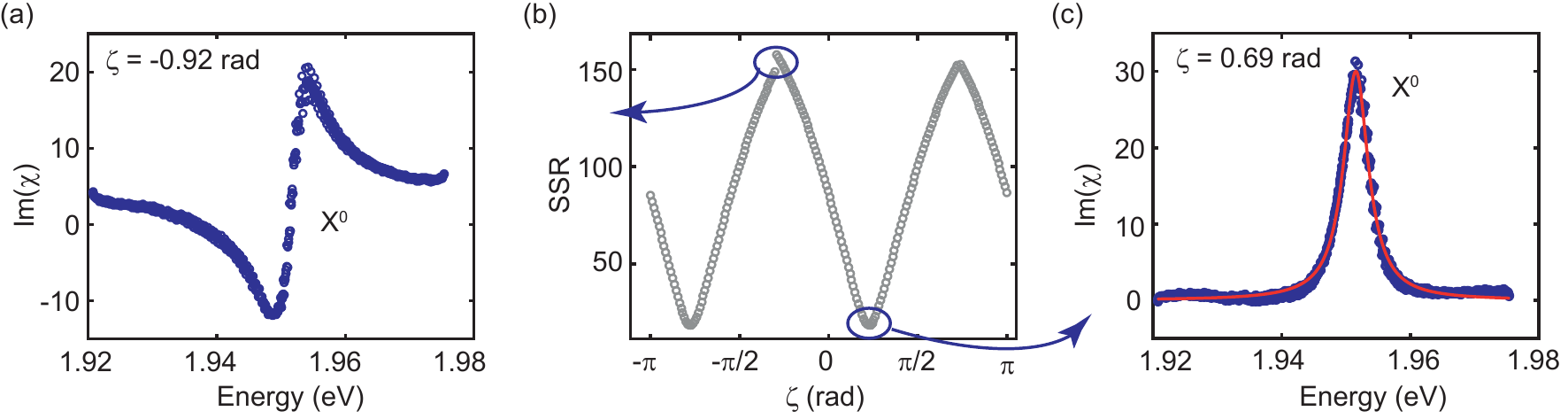}
\caption{\textbf{Kramers-Kronig and $\zeta$}. The phase factor $\zeta$ accounting for multiple thin-film interferences is varied such that the $X^0$ resonance at near zero electron density is Lorentzian. After the Kramers-Kronig transformation, the $X^0$ resonance is non-Lorentzian (a) or Lorentzian (c), depending on the value of $\zeta$.  (b) The sum of the squared residues (SSR) of a Lorentzian fit of the data is used to find the optimal $\zeta=0.69$ rad.}
\label{optimPhase}
\end{figure*}

The imaginary part of the susceptibility of a neutral exciton ($X^0$) has a Lorentzian lineshape in the absence of free carriers. A differential reflectivity spectrum at near zero electron density is transformed using the Kramers-Konig relation in Eq.~\ref{theWhole} with different values of $\zeta$ to compute the susceptibility. The integral appearing in Eq.~\ref{theWhole} is only computed over the wavelength range corresponding to the experimental spectra. The value of $\zeta$ for which the $X^0$ has a Lorentzian lineshape in the absence of carriers is then used for transforming the spectra at higher electron densities.  Fig.~\ref{optimPhase}(b) shows the sum of the  squared residues (SSR) of a Lorentzian fit of the exciton resonance. With $\zeta=0.69$ rad, the $X^0$ is well described by a Lorentzian, as in Fig.~\ref{optimPhase}(c). On the other hand, in  \ref{optimPhase}(a), at a value of $\zeta=-0.92$ rad, the  residues are maximized and the data cannot be fitted by a Lorentzian. Fig.~\ref{pic3Phases}(b) shows the differential reflectivity of $X^0$ as measured and the imaginary part of the susceptibility extracted using Eq.~\ref{theWhole}. \\

The optical susceptibility spectrum of the $X^0$ in Fig.~\ref{optimPhase}(c) is typical for our sample. The line-width of the $X^0$ in the absence of electron is extracted from a Lorentzian fit. We measure a line-width of 5~meV, close to the state-of-the-art (4~meV) observed in absorption in samples with similar structure~\cite{Cadiz2017}. As the exciton life-time was measured to be sub-picosecond~\cite{Dey2016}, the resonances observed in our susceptibility spectra are mostly homogeneously broadened, demonstrating a superior sample quality.\\

In the main text, we present absorption as a tool to non-invasively probe the 2DEG. To attest that our probe is non-invasive, we compute here the density of photo-generated electron-hole pairs and compare it to the injected electron density. 

The total power  of the LED sent on the sample is $P_{LED}=500$~nW. The LED has a broad spectrum spreading on $\Gamma_{LED}\approx 100$~meV. The power density of the LED can be estimated to be $I_{LED}=P_{LED}/\Gamma_{LED}\approx 5$~nWmeV$^{-1}$.

The main absorption of the sample occurs when $n\approx 0$. In this density regime, the absorption is dominated by the $X^0$ at an energy of $E(X^0)=1.95$~eV. We obtain an upper bound for the density of photo-generated electron-hole pairs if we assume that all the photon coming from the LED are absorbed at the $X^0$ resonance. As the $X^0$ resonance has a line-width of $\Gamma_{X^0}=5$~meV, the absorbed power is $P_{abs}=I_{LED}\Gamma_{X^0}\approx 25$~nW. The focal spot of the LED coming from a multi-mode fiber on the sample has an area of $A \approx 100$~$\mu$m$^2$. The electron-hole pair generation rate $s_{eh}$ can be therefore estimated as $s_{eh}=\frac{P_{abs}}{AE(X^0)}\approx 5\times10^{17}$~s$^{-1}$cm$^{-2}$. 

The average population of electron-hole pair is given by the product of the generation rate $s_{eh}$ with the lifetime of the electron-hole pairs  $\tau_{eh}$. The electron-hole pair lifetime in MoS$_2$ can be extracted from the homogeneous line-width obtained from four wave mixing experiments. It was measured to be $\tau_{eh}<1$~ps~\cite{Dey2016}. Using $\tau_{eh}= 1$~ps, we can estimate $n_{eh}=s_{eh}/\tau_{eh}\approx  5\times10^5$~cm$^{-2}$. As the typical electron concentrations are $n> 10^{11}$~cm$^{-2}$, $n$ is more than 5 orders of magnitude larger than $n_{eh}$, showing that our optical probe can be considered as non-invasive.\\

In order to evaluate the differential reflectivity $\frac{\Delta R}{R}$, we compare a reflectivity spectra $R$ obtained on the MoS$_2$ flake at a given gate voltage with a reference spectrum $R_0$, such that $\Delta R=R-R_0$. As the trion features in the reflectivity have a weak signal ($\frac{\Delta R}{R}\approx 5\%$), special attention need to be taken in the choice of the reference spectrum. A small interference pattern (amplitude of $\approx 5\%$) appears indeed in the raw reflectivity spectra due to reflections in the experimental setup. As they have approximatively the same amplitude as the weak signals that we want to measure, they can only be canceled by a careful choice of the reference. Ideally, $R_0$ could be acquired on the h-BN, at a position next to the MoS$_2$ flake. However, by moving the sample, the optical path length changes slightly and the interference pattern appearing in the raw reflectivity spectra does not cancel. The reflectivity spectra $R$ change significantly with varying electron density. The median of the several spectra obtained while sweeping the gate voltage  can therefore be used as the reference $R_0$ : the features appearing in  $\frac{\Delta R}{R}$ will be then only be related to changes in electron density. As $R$ and $R_0$ are all obtained at the same position, the interference pattern is canceled in $\Delta R$ and very clean differential reflectivity spectra can be acquired.

\begin{figure*}
\centering
\includegraphics[width=18cm]{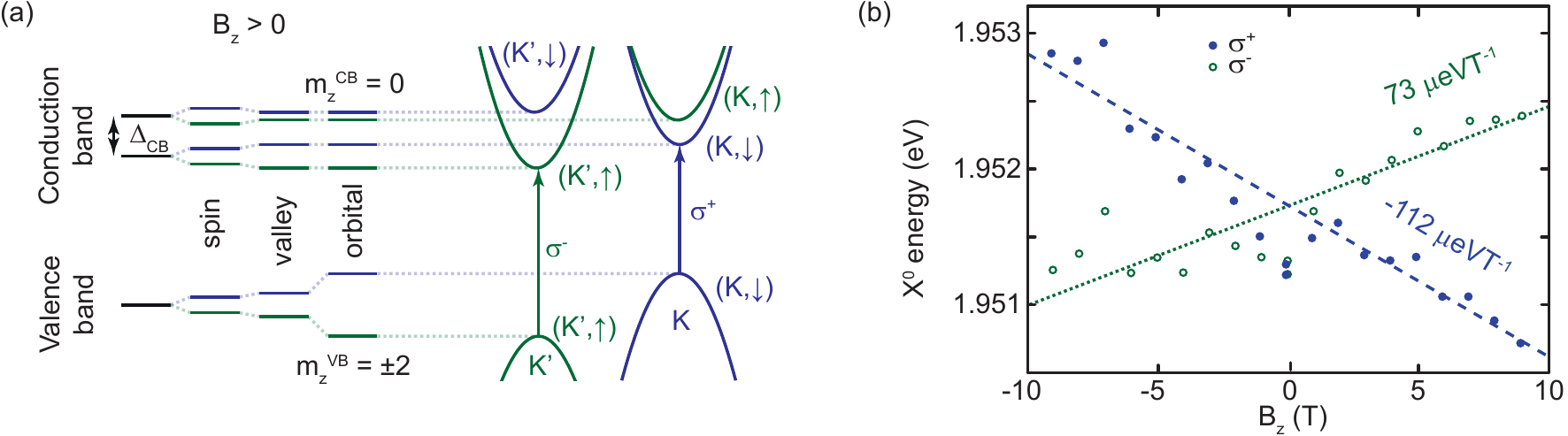}
\caption{ \textbf{Valley Zeeman effect.} (a) When increasing the magnetic field from $B_z=0$ to a finite value, the energy of the different bands changes. Valley pseudo-spin, spin and orbital momentum contribute to the energy shift of a band. (b) Energy of the $X^0$ at different magnetic fields in the two circular polarizations. The energy dependence of the exciton transition energies in both polarisations are extracted from the linear fit. In $\sigma^+$ ($\sigma^-$) polarisation, the exciton energy varies with magnetic field as -112 (73)~$\mu$eVT$^{-1}$.}
\label{X0ValleyZeeman}
\end{figure*}

\subsection{Appendix D: Selection rules and valley Zeeman effect}

For monolayer MoS$_2$, the optical response depends on the valley in which the photo-generated electron-hole pair is created. Broken inversion symmetry ensures that the electrons from the same band in the two valleys carry opposite spin, angular momentum and valley angular momentum. An out-of-plane magnetic field shifts a band in energy by a Zeeman shift $\Delta E$:
\begin{equation}
\Delta E=\underbrace{\Delta_s}_{\mathrm{spin}}+\underbrace{\Delta_\alpha}_{\mathrm{angular~momentum}}+\underbrace{\Delta_v}_{\mathrm{valley}}~~,
\end{equation} 
where $\Delta_{s}=\frac{1}{2}g_{s}\mu_Bs_zB_z$ and $\Delta_{v}=\frac{1}{2}g_{v}\mu_B\tau_zB_z$, with $g_s=1.98$\cite{Kormanyos2014} and $g_v=0.75$\cite{Kormanyos2014}. $s_z$ and $\tau_z$ are the spin and pseudo-spin (valley) operators perpendicular to the monolayer. We write $\tau_z=1$($\tau_z=-1$) for the valley pseudo-spin corresponding to the $K$ ($K'$) valley. The states at the conduction band edge are mostly made from $d_{z^2}$ orbital with a zero orbital angular momentum $m^{CB}_{z,\tau_z }=0$, yielding $\Delta^{CB}_\alpha=0$. On the other hand, the valence band is mostly composed of $d_{xy}$  and  $d_{x^2-y^2}$ orbitals with a finite orbital angular momentum $\hbar m^{VB}_{z,\tau_z}=-2\hbar\tau_z$\cite{Aivazian2015}. In the valence band, the angular momentum contribution to the Zeeman shift is finite: $\Delta^{VB}_\alpha=\mu_Bm^{VB}_{z,\tau_z}B_z$.

An optical transition must satisfy conservation of the total angular momentum, spin and momentum. The conservation laws are enforced by the selection rules that dictate which transitions can be coupled by the light field. In transition metal dichalcogenides, as the band edges are located at the $K$ and $K'$ points of the Brillouin zone, it was shown that circularly polarised light can be used to address a specific valley~\cite{Xiao2012}. A $\sigma^+$($\sigma^-$) polarised photon couples the top valence band states to  the bottom conduction band at the $K$ ($K'$) valley. 

When in $B_z>0$, the Zeeman shifts of two bands coupled via an optical transition have the same spin and valley contribution as a result of conservation rules. For an inter-band transition, only the orbital angular momentum contribution differs between the two coupled bands, as $m^{VB}_{z,\pm K}\neq m^{CB}_{z,\pm K}$. The neutral exciton ($X^0$) correspond to the coupling of the top valence band to the bottom conduction band. Its transition energy $E(X^0)$ is dictated by
\begin{equation}
E(X^0)=E_{CB}-E_{VB}-E_b(X^0)~,
\end{equation}
where $E_{CB}$ and $E_{VB}$ are the energies of states in the conduction band and valence band, respectively, and $E_b(X^0)$ is the exciton binding energy. We define $E_{CB}-E_{VB}=E_g^{B_z=0}$. In magnetic field, $E(x^0)$ is modified by the Zeeman shift of both conduction and valence band. As $s_z$ and $\tau_z$ are the same in an optical transition, the spin and valley contributions cancel:
\[
E(X^0)=E_g^{B_z=0}+(\Delta_s+\Delta_v+\Delta^{CB}_\alpha)-(\Delta_s+\Delta_v^{VB})-E_b(X^0)
\]
\begin{equation}
E(X^0)=E_g^{B_z=0}-\mu_Bm^{VB}_{z,\tau_z}B_z-E_b(X^0).
\end{equation}

As $m^{VB}_{z}$ is opposite in the two valleys, the $X^0$ energy in a finite magnetic field is different when observed in different circular polarisation of light. Fig.~\ref{X0ValleyZeeman}(a) depicts the Zeeman shifts of the different bands and shows that, when in magnetic field, the energy of the $X^0$ is different for the two circular polarisations of light. The energy difference $\Delta E(X^0)$  between $E(X^0)$ in the two valleys is given by $|\Delta E(X^0)|=2\mu_B|m^{VB}_{z,\tau_z}|B_z$.\\

The absorption energy $E(X^0)$ of the $X^0$ can be extracted from our experimental data. Fig.~\ref{X0ValleyZeeman}(b) displays $E(X^0)$  as a function of the magnetic field. It is clear that for positive magnetic fields the energy of $X^0$ increases for $\sigma^-$ polarization, while the opposite is seen for $\sigma^+$ polarization. The energetic difference $\Delta E(X^0)$ is extracted from the fits an is 185~$\mu$eVT$^{-1}$, yielding $|m^{VB}_{z,\tau_z}|=  1.6$, close to the expected value of 2, but differing from measured values on samples with similar structure~\cite{Cadiz2017}. The measurement in Fig.~\ref{X0ValleyZeeman}(b) is used to verify that $\sigma^+$ light addresses the $K$ valley.\\

We can make use of the valley Zeeman effect to verify that the optical selection rules are robust  in our experiment. In magnetic field, the $X^0$ transition energy changes in the two polarizations of light, but its lineshape  remains constant. If selection rules were not conserved in our experiment, the absorption spectrum in one polarization would contain a small amount of the features of the other polarization. A mixing of the two polarisations can be detected by computing the integrated difference between the spectra obtained in the two polarisations. We show here that up to noise level, the integrated difference between the two susceptibility spectra goes down to zero when one of the spectrum is shifted in energy by the Zeeman shift.

\begin{figure}
\centering
\includegraphics[width=80mm]{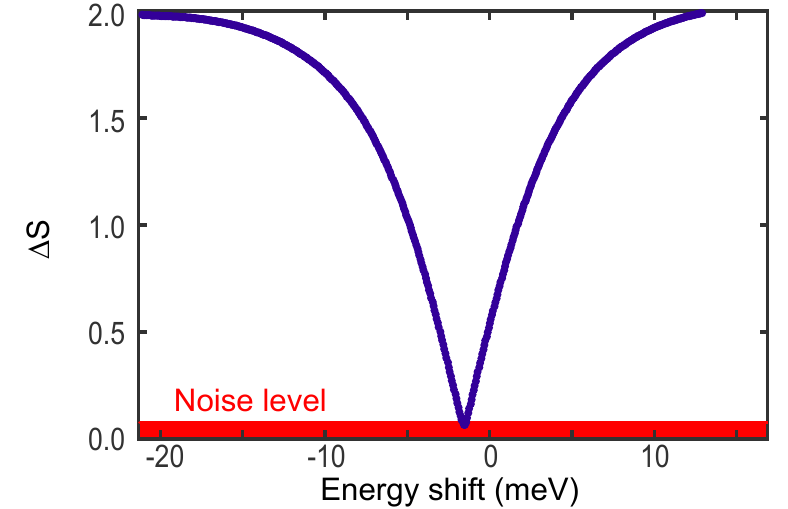}
\caption{\textbf{Robustness of optical selection rules}. The optical selection rules are verified with an accuracy reaching the noise level. $\Delta S$ is the integrated difference between the $X^0$ spectra at 9 T in the two polarizations (\textit{see} Eq.~\ref{eqDS}). The spectra are shifted in energy. The two spectra are identical when the energy shift is equal to the valley Zeeman shift.}\label{SelectionRules}
\end{figure}

The absorption signal $S^{\pm}$ in polarization $\sigma^\pm$ can be written as
\begin{equation}
S^{\pm}(E_i)=\bar{S}^{\pm}_i+N_i^{\pm}~~,
\end{equation}
where the index $i$ accounts for the discrete nature of the energy $E$ axis in the experimental spectra, $\bar{S}^{\pm}_i$ is the absorption at energy $E_i$, and $N_i^\pm$ is an independent random variable with zero mean modelling noise on the data.
To prove that the optical selection rules are conserved in our measurements, we compute the sum of the absolute value of the difference $\Delta S$ between the two spectra when they are shifted in energy by $\Delta E_j$:
\begin{equation}
\Delta S(E_j)=\sum\limits_i | S^{+}(E_i)-S^{-}(E_i+\Delta E_j) |.
\label{eqDS}
\end{equation}  
The expected value and variance of $\Delta S(E_j)$ are given by
\begin{equation}
\mathbb{E}(\Delta S(E_j))=\sum\limits_i | \bar{S}^{+}(E_i)-\bar{S}^{-}(E_i+\Delta E_j) |
\end{equation}
and
\begin{equation}
\mathrm{Var}(\Delta S(E_j))=\mathrm{Var}\left( \sum\limits_i N_i^{+} \right)+\mathrm{Var}\left( \sum\limits_i N_i^{-} \right)~~.
\end{equation}
The Central Limit Theorem implies that a sum of  random variables tends to a Gaussian random variable as the number of elements in the sum increases. Furthermore, the variance of the sum is given by the sum of the variances of the independent random variables in the sum. Here, assuming that the noise amplitude is the same at all points and in both polarizations (\textit{i.e.}  $\mathrm{Var}(N_i^+)=\mathrm{Var}(N_j^-),~\forall~i,~j$), we have 
\begin{equation}
\mathrm{Var}(\Delta S(E_j))=2h\sigma_{N}^2~~,
\end{equation}
with $\sigma_{N}$ is the standard deviation of the noise $N$ and $h$ the number of elements in the sum. $\sigma_N$ was measured on the spectra in a spectral region away from the $X^0$ resonance. Fig.~\ref{SelectionRules} shows $\Delta S$ as a function of detuning when $S^{\pm}$ are normalized such that
\begin{equation}
1=\sum\limits_i S^{\pm}(E_i)~.
\end{equation}
If the detuning is large, $\Delta S$ is 2 as the two spectra do not overlap at all. When the detuning reaches the value of the Zeeman shift, $\Delta S$ should reach zero. However, noise prevents reaching a zero value. The red domain in Fig. \ref{SelectionRules} shows possible values of $\Delta	 S$ taking noise into account, in the case of perfect selection rules. As the minimal value of the measured $\Delta	 S$ (blue curve) falls in the noise level (red domain) at the value of the Zeeman shift, we can conclude that selection rules are conserved to the detection limit of our setup.

\subsection{Appendix E: Theory of trion absorption in a 2DEG}
\label{zetheory}

\begin{figure*}
\includegraphics[width=180mm]{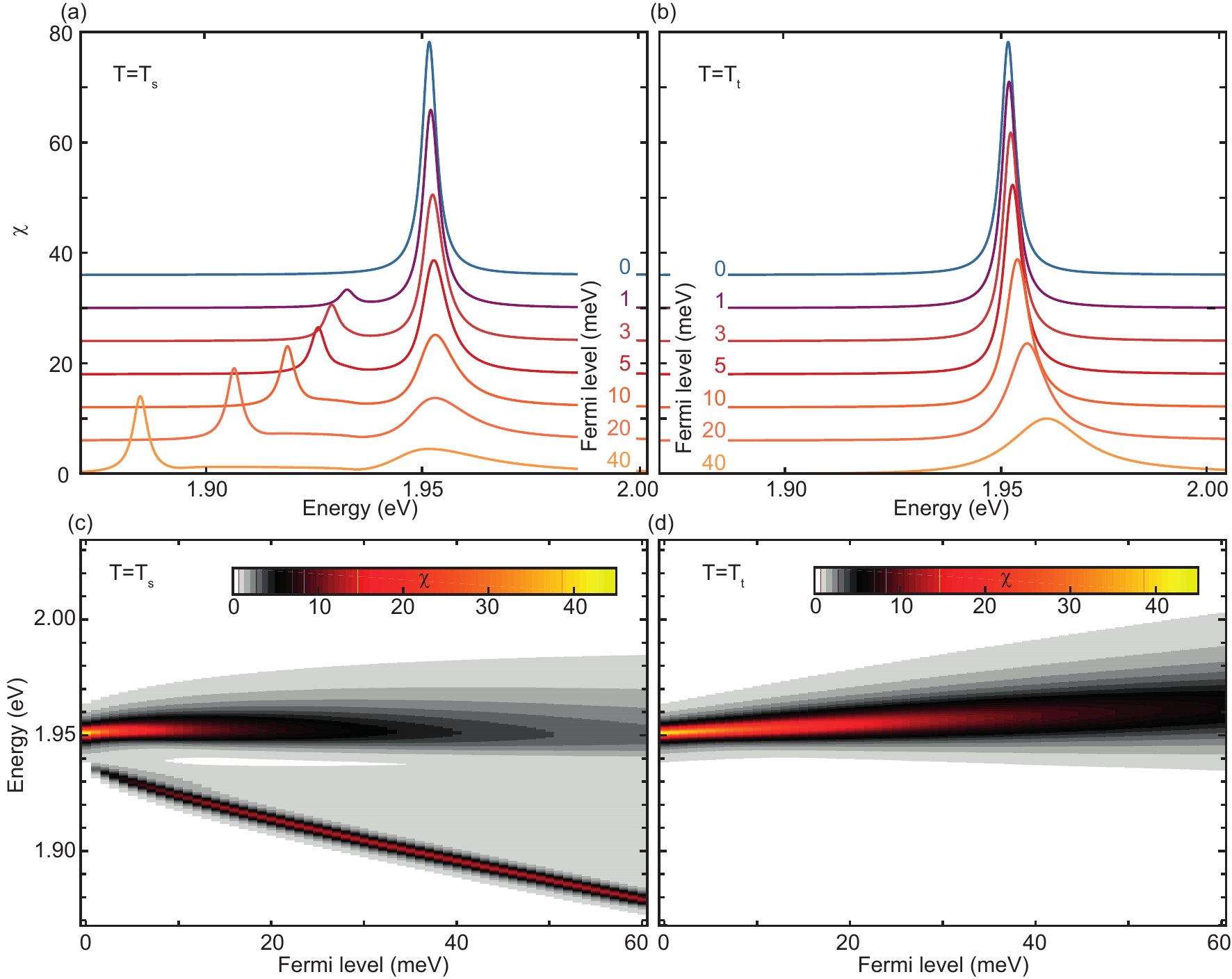}
\caption{\textbf{Suris model of 2DEG optical susceptibility.} Optical susceptibility of an exciton simulated following Eq.~\ref{susc}. The exciton interact with a fully spin-polarised Fermi sea. (a) The electron of the photo-generated excitons have the same spin as the electrons of the Fermi sea. Excitons experience a repulsive interaction with electrons (triplet collisions). The $X^0$ peak becomes more asymmetrical toward the high energy side with higher electron density. (b) The electron of the photo-generated excitons have opposite spin to that of the electrons of the Fermi sea. Excitons interact attractively with electrons (singlet collisions). A low energy peak emerges around 1.9~eV as consequence of interactions, while the $X^0$ stays at 1.95~eV. (c) Optical susceptibility in the case of singlet collision as a colormap at different Fermi levels and different photon energies. (d) Same as in (c) but for triplet collisions. The simulations presented here were obtained using $m^*_{CB}=0.44$\cite{Kormanyos2014}, $m^*_{VB}=0.5$~\cite{Rybkovskiy2017}, $E_b(X^0)=260$~meV, $E_b(X^-)=17$~meV, $E(X^0)=1.952$~eV, $\alpha=1$, and $\gamma=2.0$~meV.    }\label{suris}
\end{figure*}

When the Fermi level $E_F$ is of the same order of magnitude as the trion binding energy $E_b(X^-)$ and less than the exciton binding energy $E_b(X^0)$, the description of the trion as a three particle body is no longer valid~\cite{Efimkin2017,Suris2001}. In this limit, trions appear as a consequence of the exciton - Fermi sea interaction: the exciton energy splits in two branches when it interacts with a Fermi sea, forming exitons and trions. In this picture, it is more accurate to name the two branches in terms of Fermi polarons~\cite{Massignan2014}. The upper energy branch, the exciton, is then the repulsive polaron, while the trion forming the lower energy branch is referred to as the repulsive polaron~\cite{Massignan2014}.

\begin{figure}
\centering
\includegraphics[width=80mm]{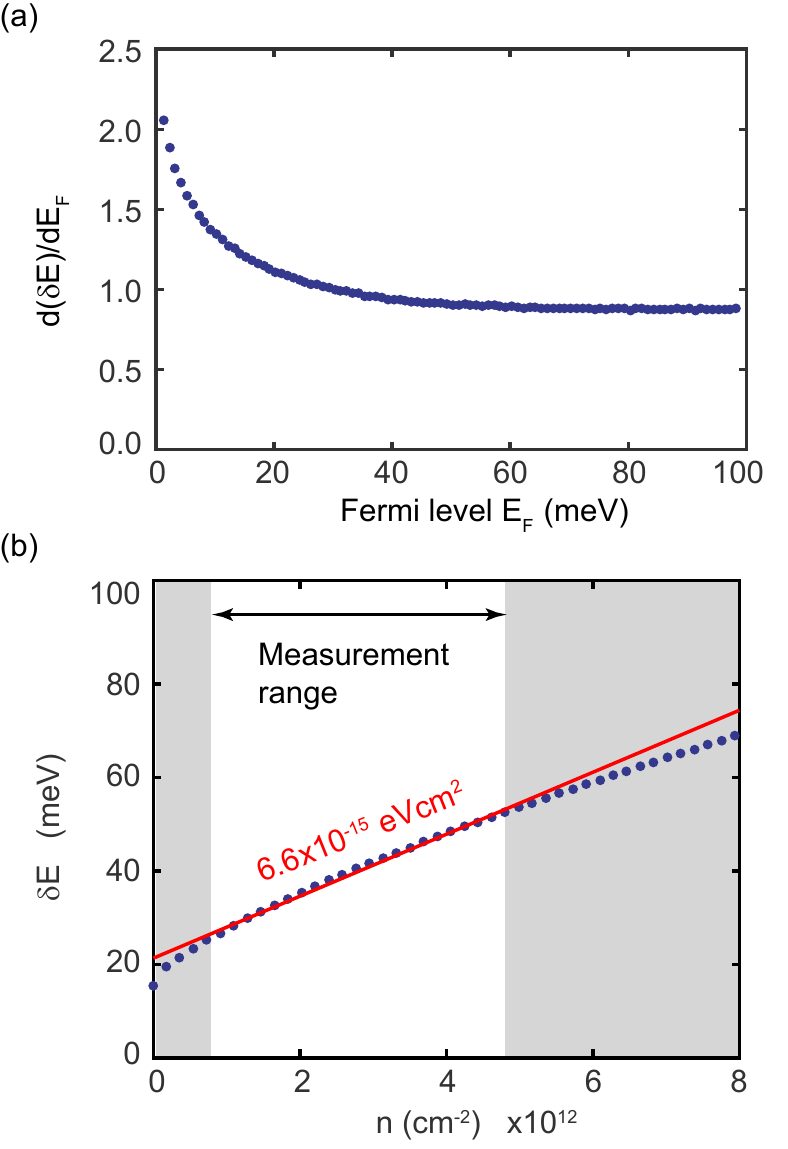}
\caption{\textbf{Optics as a measure of the Fermi level}. The susceptibility in Eq.~\ref{susc} can be computed at various Fermi-level, as in Fig.~\ref{suris}. From the simulated spectra we can extract the energy of the $X^0$ and of the $X^-$. (a) Gradient of $\delta E$ as a function of the Fermi level. At high Fermi level, the gradient tends to a value of 1. (b) $\delta E$ as a function of the carrier density. The red solid line is a linear fit of the simulated $\delta E$ in the range of densities where trion and exciton can be experimentally resolved in the susceptibility spectra. The parameters used for this simulations are the same as those used in Fig.~\ref{suris}.}
\label{SI_SurisSlope}
\end{figure}

Photo-generated excitons interact with the Fermi-sea in two ways: either an exciton captures an electron and creates an $X^-$ (attractive interaction) or  an electron scatters off an exciton (repulsive interaction). These two interactions contributes to the self-energy ($\Xi$) in the 2D exciton optical susceptibility~\cite{Efimkin2017,Suris2001,HandK}.
\begin{equation}
\chi(\hbar\omega)=-2|d_{cv}|^2\dfrac{|\psi(\mathbf{r}=0)|^2}{\hbar\omega+i\gamma-E(X^0)-\Xi}~~, 
\end{equation}
where $E(X^0)$ is the exciton energy, $\gamma$ accounts for broadening, $d_{cv}$ is the intervalley optical dipole moment and $\psi$ is the exciton wavefunction.

It was shown that in the regime when $E_F<E_b(X^-)$, the self energy $\Xi$ can be  written as~\cite{Efimkin2017}
\begin{equation}
\Xi(\hbar\omega)=\int\limits_0^{\infty}g_{2D}f_{\textrm{FD}}(\epsilon)T(\hbar\omega+\epsilon)\mathrm{d}\epsilon~~,\label{susc}
\end{equation}
where $f_{\textrm{FD}}(\epsilon)$ is the Fermi-Dirac distribution describing the occupation of the Fermi-sea, $g_{2D}=\frac{m^*_{CB}}{2\pi\hbar^2}$ is the two-dimensional density of states (without spin-degeneracy), $m^*_{CB}$ is the effective electron mass in the conduction band and $T(\epsilon)$ is the two-particle $T$-matrix. 

Under the assumption that the electrons of a Fermi-sea interact only with excitons composed with an electron of opposite spin (singlet collision) forming a bound state (singlet trion) the $T$-matrix elements are~\cite{Efimkin2017}
\begin{equation}
T(\hbar\omega)=T_{s}(\hbar\omega)=\frac{2\pi\hbar^2}{\mu_T}\frac{1}{\ln\left(\dfrac{-E_b(X^-)}{\hbar\omega -E(X^0)+i\gamma }\right)} ~,\label{mdo}
\end{equation}
where $\mu_{T}$ is the reduced exciton-electron mass. $\mu_{T}$ can be written as
\begin{equation}
\frac{1}{\mu_{T}}=\frac{1}{m^*_{CB}}+\frac{1}{m^*_{VB}+m^*_{CB}}~,
\end{equation}
with $m^*_{VB}$ the effective electron masses in the  valence band.\\

Suris~\cite{Suris2001} derived a model of absorption of a 2DEG considering the effect of both singlet and triplet collisions. In a singlet collision, exciton-electron interaction has a bound state corresponding to the singlet trion. On the contrary, in triplet collision, they are no bound states of the exciton-electron interaction. The absence of a bound state comes the fact that triplet trions are unbound in the absence of magnetic fields~\cite{Whittaker1997}.

In his model,~\cite{Suris2001} the two types of interactions are introduced by decomposing the $T$-matrix in two parts $T=\frac{1}{2}T_{s}+\frac{3}{2}T_{t}$, with $T_{s}$ accounting for singlet interactions and $T_{t}$ for triplet interactions in the scattering of an electron on an exciton. As his scattering matrix elements $T_{s}$ are in agreements with those of Eq.~\ref{mdo}, we generalize here the results from Ref.~\cite{Efimkin2017} to the case of singlet and triplet interactions inspired by the work from Suris~\cite{Suris2001}.

The $T$-matrix elements accounting for triplet interactions can be written as~\cite{Suris2001}
\begin{equation}
T_{t}(\hbar\omega)=\frac{2\pi\hbar^2}{\mu_T}\frac{1}{\ln\left(\dfrac{-E_b(X^0)}{\hbar\omega -E(X^0)+i\gamma }\dfrac{E_b(X^0)}{\alpha E_b(X^-)}\right)}~~, \label{Tt}
\end{equation}
where $\alpha\approx 1$ is a number.

Fig.~\ref{suris} shows the contributions of the different $T$-matrix components to the optical susceptibility. Fig.~\ref{suris} (a, c) show the optical susceptibility resulting only from singlet scattering ($T=T_{s}$). A sharp peak around 1.9~eV emerges with a finite Fermi level. It corresponds to the $X^-$, or more precisely to the attractive polaron. The $X^0$ resonance at 1.95~eV (repulsive polaron) is also influenced by the interaction. The $X^0$ resonance looses indeed quickly in amplitude and the line-shape becomes more asymmetrical as the Fermi-level increases.

When only triplet scattering ($T=T_{t}$) is allowed, as in Fig.~\ref{suris}(b, d), there is no low energy peak, as they are no bound triplet trions. The $X^0$ resonance is however still modified by the presence of the Fermi sea. The amplitude of the $X^0$ decreases and the $X^0$ shifts to higher energy. The line-shape of the $X^0$ is also modified by the presence of electrons: as the Fermi level increases, the $X^0$ peak becomes less and less Lorentzian as a tail grows on its high energy side.

Comparison between Fig.~\ref{suris} (a) and Fig.~\ref{suris} (b) informs that an attractive electron-exciton interaction is responsible for a sharp low energy resonance (the $X^-$ or attractive polaron), while a repulsive electron-exciton interaction tends to create a high energy tail to the $X^0$. The prediction of the polaron model is in good agreement with our experimental data presented in Fig. 2 of the main text. However, the high energy trion $X^-_{HE}$ has its line-width increase with increasing electron density due to the different nature of the Coulomb interaction.

The energetic difference between neutral exciton and trion, $\delta E=E(X^0)-E(X^-)$ can be used to measure variations of the Fermi level~\cite{Huard2000}. Hawrylak~\cite{Hawrylak1991} showed that the $X^-$ is the ground state of an electron-hole pair in a Fermi sea. The $X^0$ is then an ionised $X^-$. In the ionisation process, the electron must be dragged to the first available state in the conduction band, at the Fermi level. The energy cost of ionising an $X^-$ is therefore increasing with the Fermi-level, pushing the $X^0$ toward higher energies. In this picture, $\delta E$ varies as the Fermi level $E_F$, \textit{i. e.} $\frac{d}{dE_F}\delta E=1$. 

In the framework of exciton-polarons, changes in  $\delta E$ can also be related to changes in Fermi level. Fig.~\ref{SI_SurisSlope}(a) shows the evolution of the gradient $\frac{d}{dE_F}\delta E$ as a function of the Fermi level extracted from simulations, as in Fig.~\ref{suris}. In the high density regime, the slope $\frac{d}{dE_F}\delta E=1$ in agreements with Ref.~\cite{Hawrylak1991}. However, as the electron density decreases, the slope increases up to $\frac{d}{dE_F}\delta E=2$.

In our experiment, we can observe trions at Fermi levels ranging from 5~meV to 25~meV (two bands filled). In this regime, the gradient is varying slightly from 1.5 to 1, and equals on average 1.2. Fig.~\ref{SI_SurisSlope}(b) shows $\delta E$ as a function of the electron density $n$. We find that in the regime where we can observe the trion, $\delta E$ varies as $6.6\times 10^{-15}$~eVcm$^{2}$ with $n$, using an electron mass $m_{CB}^*=0.44$\cite{Kormanyos2014}. 

The gradients $\frac{d}{dE_F}\delta E$ measured experimentally presented in the main text for $X^-_{LE}$($X^-_{HE}$)are $6.1\times 10^{-15}$~eVcm$^{-2}$ ($6.3\times 10^{-15}$~eVcm$^{-2}$) in a 9~T magnetic field and $5.6\times 10^{-15}$~eVcm$^{-2}$ ($5.9\times 10^{-15}$~eVcm$^{-2}$) in the absence of a magnetic field. These values are in good agreement with the slope $\frac{d}{dE_F}\delta E=6.6\times 10^{-15}$~eVcm$^{-2}$ extracted from our simulations, as shown in Fig.~\ref{SI_SurisSlope}(b). In this rather limited range of electron densities, we cannot observe a significant change of the slope $\frac{d}{dE_F}\delta E$.

\subsection{Appendix F: Band filling and optics in M{o}S$_2$}

\begin{figure*}
\centering
\includegraphics[width=180mm]{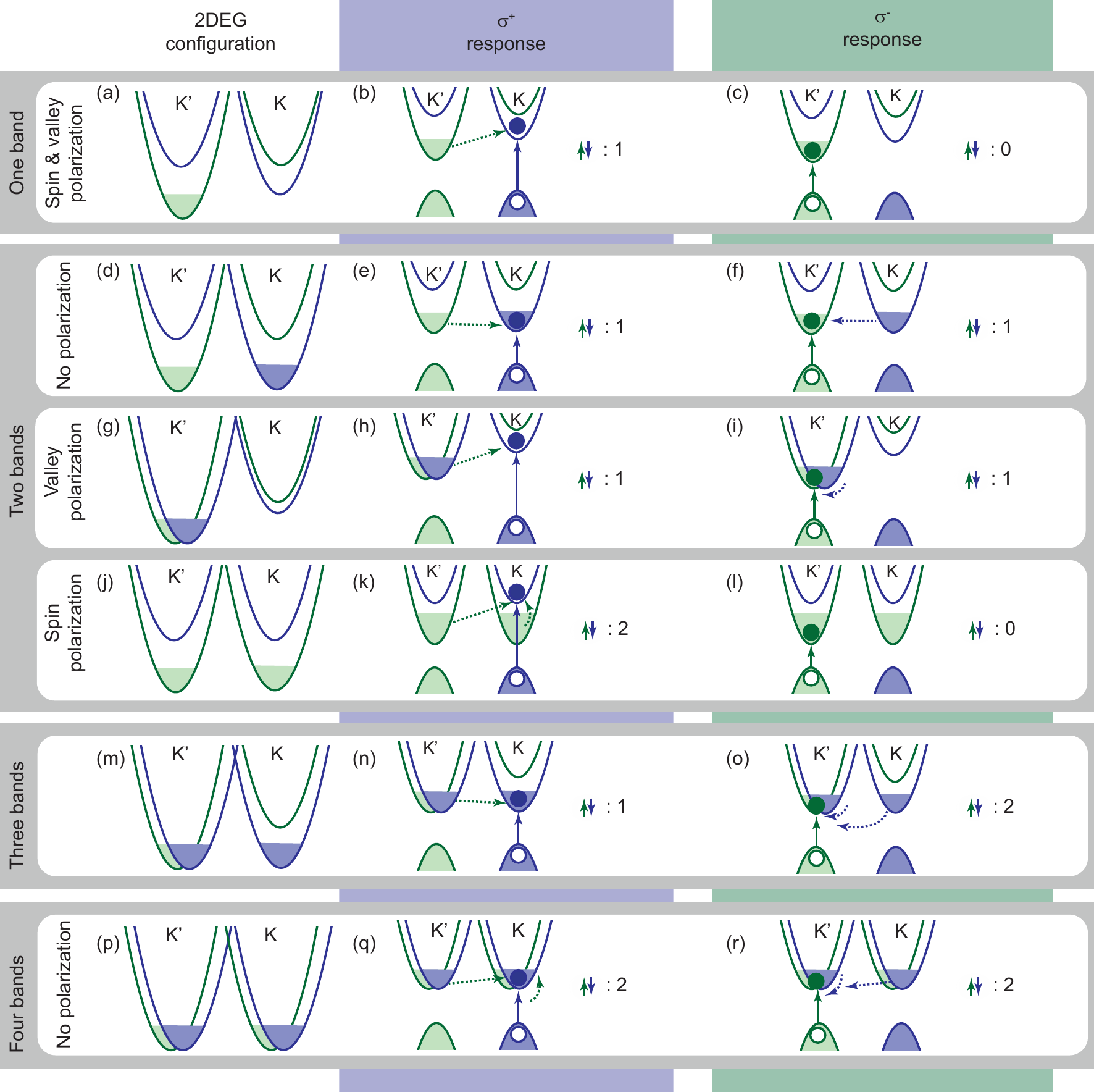}
\caption{\textbf{Possible band filling in MoS$_2$ and consequences on the optical susceptibility}.}
\label{allPossi}
\end{figure*}

In the main text, we explain our optical spectra by a spin polarization of the 2DEG electrons. This section aims at describing how only a spin polarization of the 2DEG with two bands filled can explain the optical susceptibilities presented in the main text in Fig.~2. The optical susceptibility of MoS$_2$ in a magnetic field is dominated by two peaks at lower energy than the neutral exciton ($X^0$) in one polarization. In the other polarization, the optical susceptibility is dominated by the neutral exciton and no other resonances appear in the susceptibility before the appearance of the $Q$-peak. We review here the different ways of filling the bands and their consequence on the optics by commenting Fig.~\ref{allPossi}.\\

\paragraph*{One band filled.} When only one band is filled the 2DEG is simultaneously spin- and valley-polarized. Such a band filling, as in Fig.~\ref{allPossi}(a) is observed in MoSe$_2$ when placed in a magnetic field~\cite{Back2017}. It was observed~\cite{Back2017} that when an electron-hole pair is created in the filled band, only the neutral exciton ($X^0$) appears in the optical susceptibility (Fig.~\ref{allPossi}(c)). On the other hand, when the electron-hole pair is created in the empty band with opposite spin as in Fig.~\ref{allPossi}(b), a low energy resonance identified as a trion or, more accurately, an attractive polaron dominates the susceptibility. This result is completely compatible with theory presented in Section~\ref{zetheory} : an attractive polaron (low energy resonance) is formed when the spin of the electrons in the 2DEG is opposite to that of the electron component of the photo-generated electron-hole pairs. As we find that we have two types of low energy resonances, we can conclude that we fill more than a single band.\\

\paragraph*{Two bands filled.} If the two spin-split conduction bands were energetically well far apart, in the absence of many-body interactions, it would be expected that only the two lowest energy bands would be populated, as in Fig.~\ref{allPossi}(d). In this case, the 2DEG filling is symmetric in both spin and valley population. The optical response is therefore also expected to be independent on the polarization of the light, in contradiction to our measurements.

Another possibility of two bands filling is to create a valley polarization: all electrons are located at the $K'$ of the Brillouin zone with both spin up and spin down population, as depicted in Fig.~\ref{allPossi}(g). When an electron-hole pair is formed in the same valley $K'$ than the 2DEG electrons (Fig.~\ref{allPossi}(i)), the optical response should be similar to that of two dimensional structures of conventional direct band-gap semiconductor with band edges at the $\Gamma$-point, such as GaAs. In the absence of magnetic field, in GaAs, only the singlet trion is observed~\cite{Shields1995}. As the magnetic field increases, the triplet trion emerges as a low energy shoulder of the $X^0$, before being resolved as a separated resonance at high magnetic field.  In our case, it would then be expected that when an electron-hole pair is created in the $K'$ valley ($\sigma^-$ polarized light), as in Fig.~\ref{allPossi}(i), where the 2DEG electrons are, at least one resonance should appear in the optical susceptibility. When measuring the susceptibility using $\sigma^+$ polarized light (Fig.~\ref{allPossi}(h)), there is also one possibility of creating a singlet trion. If the 2DEG was forming a valley polarization in MoS$_2$ with two filled bands, we would then have at least on resonance in both polarizations. This is also in contradiction with our measurements.

As written in the main text, our results are best explained by a spin polarization (Fig.~\ref{allPossi}(j)): when an electron-hole pair is formed in the $K$ valley ($\sigma^+$ polarized light), the promoted electron carries opposite spin to that of the electrons, as shown in Fig.~\ref{allPossi}(k). Two possible singlet trions can be observed in this polarization. On the other hand, when a $\sigma^-$ polarized photon promotes an electron with spin up (Fig.~\ref{allPossi}(l)) it can only interact repulsively with electrons of the same spin, as explained in Section~\ref{zetheory}. No bound states are then observed in this polarization, in agreements with our experimental results.\\

\paragraph*{Three bands filled.} The experimental result obtained on MoSe$_2$ tell us that the optical response of an electron-hole pair created in a filled band is barely modified by the presence of the electrons~\cite{Back2017}. Merely a decrease in oscillator strength and an energetic shift of the $X^0$ resonance are indeed observed. The three bands filling case as represented in Fig.~\ref{allPossi}(m), when probed with $\sigma^+$ light (Fig.~\ref{allPossi}(n)) is therefore similar to the case of the two-bands valley polarization in Fig.~\ref{allPossi}(h). At least one resonance should then be observed in this polarization. When the light is $\sigma^-$ polarized (Fig.~\ref{allPossi}(o)), the three band case is then similar to the two-band spin-polarized case in Fig.~\ref{allPossi}(k) and at least two resonances should be observed in this polarization. The three band-filling case is therefore in contradiction with our measurements.\\

\paragraph*{Four bands filled.} Similarly to the unpolarized two-band case, as thefour bands 2DEG band filling, as in Fig.~\ref{allPossi}(p), is symmetric in both spin and valley filling, the optical response should also be similar in both polarizations.

\end{document}